\documentclass[12pt]{article}
\usepackage[numbers]{natbib}
\usepackage{a4}
\usepackage{amsmath}
\usepackage{amssymb}
\usepackage{latexsym}
\usepackage{amssymb}
\usepackage{epsfig}
\usepackage{natbib} 
\def \TP{{\mathrm{P}}}

\def \pd{\partial}

\begin{document}
\title{{\bf Dislocation Field Theory in 2D:\\ 
Application to Graphene}}
\author{
Markus Lazar~$^\text{a,b,}$\footnote{
{\it E-mail address:} lazar@fkp.tu-darmstadt.de (M.~Lazar).
}
\\ \\
${}^\text{a}$ 
        Heisenberg Research Group,\\
        Department of Physics,\\
        Darmstadt University of Technology,\\
        Hochschulstr. 6,\\      
        D-64289 Darmstadt, Germany\\
${}^\text{b}$ 
Department of Physics,\\
Michigan Technological University,\\
Houghton, MI 49931, USA
}

\date{\today}    
\maketitle

\begin{abstract}
A two-dimensional (2D) dislocation continuum theory is being introduced.
The present theory adds elastic rotation, dislocation density,  
and background stress to the classical energy density of elasticity.
This theory contains four material moduli.
Two characteristic length scales are defined in terms of the four 
material moduli.
Non-singular solutions of the stresses and elastic distortions
of an edge dislocation are calculated. 
It has been 
pointed out that the elastic strain agrees well with experimental data
found recently for an edge dislocation in graphene. 
\\

\end{abstract}

\newpage

\section{Introduction}

A challenging and active research field is the investigation of
the material behaviour of graphene, especially the study of dislocations 
in graphene (see, e.g.,~\citep{Zak,Yav,Chen,Yov,Juan}). 
Graphene is a two-dimensional (2D) material with extraordinary physical 
properties.
In a recent experiment~\citep{Warner}, the elastic strain and rotation fields
produced by an edge dislocation in graphene have been observed 
for the first time. 
It was reported that the lattice rotation is quite appreciable at 
the dislocation core. Also it was noted that the measured elastic 
strain contours, do not agree with the corresponding contours calculated 
in classical elasticity theory.
This indicates that a general dislocation continuum theory including
the elastic rotation, is needed for a theoretical prediction of realistic
strain and rotation contours.
Dislocations are 
critical for understanding plasticity in 2D crystals and predicting mechanical properties.
Dislocations are the fundamental carrier of plasticity of materials and 
little is known about their effect in 2D crystals.

This paper shows that the so-called dislocation field or dislocation
gauge theory (see, e.g.,~\citep{KE,EL,Lazar00,LA08,LA09,Lazar09,LH10,AL10})
is a promising and excellent candidate to fulfill the
requirements mentioned above, 
and to give contours which agree with experimental data.
In~\citep{LA09} a dislocation field theory, which can be considered as the
dislocation gauge theory of the three-dimensional translation group $T(3)$
was developed.
The idea of a static dislocation field theory, is to use
three terms in the general distortion energy density. 
One term contains the elastic strain and the elastic rotation fields.
Another one proportional to the dislocation density tensor 
having the meaning of dislocation core energy density and 
a term containing a background stress tensor, which
is needed for self-equilibrating of the dislocations. 
It is important to mention that the force stress tensor is not 
symmetric anymore.
In~\citep{LA09}, non-singular solutions for screw and edge dislocations
were found.
This paper adopts the framework of~\citep{LA09} 
in order to formulate a dislocation field theory for two-dimensional
materials,
which is a gauge theory of the two-dimensional translation group
$T(2)$. We suggest using such a dislocation field theory 
as a 2D dislocation continuum theory for dislocations in graphene.
We propose a 2D dislocation field theory, 
because the strain fields around dislocations differ from 
those given by classical elasticity with line singularities.

The outline of this paper is as follows.
In Section~2, the fundamental framework of 2D dislocation continuum
field theory is presented.
In Section~3, the non-singular solutions of the stress and elastic
distortion fields are given. 
In addition, the components of the dislocation density 
vector and the effective Burgers vector are calculated. The physical
features of the obtained solutions are presented in suitable plots.
Section~4, concludes our work.

\section{Basic Framework}
In 2D a dislocation is characterized by the Burgers vector which 
can be in $x$- and $y$-directions with components $b_x$ and $b_y$. 
There is no $z$-direction in 2D. For that reason a dislocation in 2D is a point
dislocation. 
The dislocation is located at the dislocation point.
In real 2D materials only edge dislocations are possible since 
the Burgers vector
is constrained to lie in the $xy$-plane.
The two physical state quantities in the static theory 
of dislocations are the elastic distortion tensor
\begin{align}
\label{B}
\beta_{ij}=u_{i,j}-\beta^\TP_{ij}\,, \qquad i,j =x,y
\end{align}
and the dislocation density vector
\begin{align}
\label{DD}
\alpha_{i}&=\epsilon_{kl}\beta_{il,k}\,,\\
\alpha_{i}&=-\epsilon_{kl}\beta^\TP_{il,k}\,,
\end{align}
which is the measure how much the elastic distortion tensor 
$\beta_{ij}$ and the plastic distortion tensor $\beta^\TP_{ij}$ 
are incompatible. 
$\epsilon_{ij}=-\epsilon_{ji}$, $\epsilon_{xy}=1$ is the totally
antisymmetric second rank tensor.
The displacement vector is denoted by $u_i$ and is not
a physical state quantity.
Since in 2D a dislocation is a point dislocation, 
there is no Bianchi identity for dislocations unlike 3D
where a dislocation is a line defect.
Also it holds
\begin{align}
\label{T}
T_{ijk}=\epsilon_{jk}\alpha_i=\beta_{ik,j}-\beta_{ij,k}\,,\qquad\qquad
\alpha_i=\frac{1}{2}\, \epsilon_{jk}T_{ijk}\,,
\end{align}
where $T_{ijk}$ is Cartan's torsion tensor in 2D (see, e.g., \citep{Mielke}).

The deformation energy density consists of three pieces
\begin{align}
\label{W}
W=W_{\text{el}}+W_{\text{di}}-W_{\text{bg}}\,.
\end{align}
The first piece is the elastic distortion energy density
\begin{align}
W_{\text{el}}=\frac{1}{2}\,\sigma_{ij}\beta_{ij}\,,
\end{align}
the second piece is the dislocation energy density
\begin{align}
\label{W-core}
 W_{\rm di}=
\frac{1}{2}\, H_i \alpha_{i}\,,
\end{align}
playing the role of the dislocation core density 
and finally,
the third piece is the background part
\begin{align}
\label{L-null}
W_{\rm bg}=\sigma^0_{ij} \beta_{ij}\,,
\end{align}
containing the contribution of the residual 
or background stress tensor $\sigma^0_{ij}$, fulfilling the condition
$\sigma^0_{ij,j}=0$,
needed to equilibrate dislocations.

The specific response fields in the framework of dislocation field theory
shall be given for an isotropic, linearly elastic medium.
The force stress tensor is defined by
\begin{align}
\label{sigma}
\sigma_{ij}=\frac{\pd W_{\text{el}}}{\pd \beta_{ij}}
=\lambda\, \delta_{ij}\beta_{kk}+2\mu\, \beta_{(ij)}+2\gamma\, \beta_{[ij]}\,,
\end{align}
where the symmetric part, 
$\beta_{(ij)}=(\beta_{ij}+\beta_{ji})/2$, is the elastic strain tensor and
the skew-symmetric part,
$\beta_{[ij]}=(\beta_{ij}-\beta_{ji})/2$,  determines
the elastic rotation.
Here $\mu$ and $\lambda$ are the Lam{\'e} coefficients.
The coefficient $\gamma$ is an additional material parameter 
due to the skew-symmetric part of the elastic distortion 
(the elastic rotation). 
Thus, $\gamma$ is the modulus of rotation (see also~\citep{LA09}).
The skew-symmetric stress $\sigma_{[ij]}$ is caused by the (local) elastic
distortion $\beta_{[ij]}$.
In 2D the trace of the elastic distortion tensor is
\begin{align}
\sigma_{kk}=\sigma_{xx}+\sigma_{yy}=2(\lambda+\mu)\,\beta_{kk}\,,
\end{align}
due to $\delta_{kk}=2$.
The response to the dislocation density vector is given by
\begin{align}
\label{H}
H_{i}=\frac{\pd{W}_{\rm di}}{\pd \alpha_{i}}
=c\, \alpha_i
\end{align}
and is the dislocation excitation vector. 
In 2D, the dislocation density vector is already irreducible with respect to
the two-dimensional group of isotropy, $SO(2)$, and it possesses two
independent vector components $(\alpha_x,\alpha_y)$ (see, e.g.,~\citep{Mielke,Hehl95}). 
$c$ is the dislocation modulus.
$H_i$ has the physical meaning of a pseudo-moment stress vector 
(see also,~\citep{LA09}).
Since in 2D we have a dislocation density vector, the theory possesses only
one dislocation modulus unlike 3D where three dislocation moduli 
are present.
This 2D dislocation continuum field theory contains
four material constants: the two Lam\'e moduli $\mu$ and $\lambda$,
the rotation modulus $\gamma$, and the dislocation modulus $c$.
The positive semi-definiteness of $W$, $W\ge 0$, requires the restriction
\begin{align}
\label{PSD1}
\mu\ge 0,\qquad\gamma \ge 0,\qquad \mu+\lambda\ge 0\,,\qquad c\ge 0\, .
\end{align}

The Euler-Lagrange equations of $W$ with respect to the elastic distortion
tensor are given by
\begin{align}
\frac{\delta W}{\delta\beta_{ij}}=
\frac{\pd W}{\pd \beta_{ij}}- 
\pd_k \frac{\pd W}{\pd\beta_{ij,k}}=0\,,
\end{align}
which give the fundamental field equations for dislocations,
the dislocation equilibrium condition.
They read in terms of the response quantities
\begin{align}
\label{ME-SF}
\epsilon_{jk}H_{i,k}+\sigma_{ij}=\sigma^0_{ij}
\end{align}
and with Eq.~(\ref{H}) 
\begin{align}
\label{ME-SF2}
c\, \epsilon_{jk}\alpha_{i,k}+\sigma_{ij}=\sigma^0_{ij}\,.
\end{align}
Differentiating Eq.~(\ref{ME-SF}) with respect to $x_j$,
the force equilibrium condition of the force stress tensor $\sigma_{ij}$
follows
\begin{align}
\label{FE}
\sigma_{ij,j}&= 0\,.
\end{align}

By means of Eq.~(\ref{DD}), Eq.~(\ref{ME-SF2}) takes the following form
\begin{align}
\label{ME2}
&c\,(\beta_{ik,jk}- \beta_{ij,kk}) +\sigma_{ij}=\sigma^0_{ij}.
\end{align}
Using the inverse constitutive relation for $\beta_{ij}$
\begin{align}
\label{CR-B}
\beta_{ij}= \frac{\gamma + \mu}{4\mu\gamma}\,\sigma_{ij} +
\frac{\gamma - \mu}{4\mu\gamma}\,\sigma_{ji}
 - \frac{\nu}{2\mu (1 + \nu)}\,\delta_{ij}\, \sigma_{kk}\,,
\end{align}
where the 2D Poisson ratio $\nu$ is expressed
in terms of the Lam{\'e} coefficients
\begin{align}
\label{}
\nu=\frac{\lambda}{2\mu +\lambda}\,,\qquad
\lambda=\frac{2\mu\nu}{1-\nu}
\end{align}
and the trace of the elastic distortion tensor, 
which gives the elastic dilatation,
\begin{align}
\beta_{kk}=\beta_{xx}+\beta_{yy}=\frac{1-\nu}{2\mu(1+\nu)}\, \sigma_{kk}\,,
\end{align}
and Eq.~(\ref{FE}), the field equation~(\ref{ME2}) can be rewritten in 
terms of the force stress tensor.
The result reads
\begin{align}
\label{ME3}
 & c \Big[
\frac{\gamma-\mu}{4\mu\gamma}\, \sigma_{ki,jk}
-\frac{\nu}{2\mu(1+\nu)}\, \sigma_{kk,ij}
-\frac{\gamma+\mu}{4\mu\gamma}\, \sigma_{ij,kk}
-\frac{\gamma-\mu}{4\mu\gamma}\, \sigma_{ji,kk}
+\frac{\nu}{2\mu(1+\nu)}\, \delta_{ij}\sigma_{ll,kk}\Big]
+\sigma_{ij}=\sigma^0_{ij}\,.
\end{align}
Eq.~(\ref{ME3}) is the fundamental field equation 
for dislocations in terms of the force stress tensor
derived in the framework of dislocation field theory in 2D. 
Thus, Eq.~(\ref{ME3}) is the equation of motion for the stress tensor.
Eq.~(\ref{ME3}) will serve exciting solutions for the 
dislocation fields.

It can be seen in Eq.~(\ref{ME3}) that the components of 
the force stress tensor $\sigma_{ij}$ are coupled in that equation. 
We can construct two simple, uncoupled, inhomogeneous Helmholtz equations 
for the trace $\sigma_{kk}$, and the skew-symmetric part $\sigma_{[xy]}$.
The trace of Eq.~(\ref{ME3}) gives
\begin{align}
\label{ME-dil}
\Big[1-\frac{c}{2\mu(1+\nu)}\, \Delta\Big]\sigma_{kk}=\sigma^0_{kk}\,,
\end{align}
where $\Delta$ denotes the 2D Laplacian. From the skew-symmetric part 
of Eqs.~(\ref{ME-SF2}) and (\ref{ME3}) and some simple algebra, 
we obtain
\begin{align}
\label{ME-rot}
\Big[1-\frac{c(\mu+\gamma)}{4\mu\gamma}\, \Delta\Big]\sigma_{[xy]}
=\sigma^0_{[xy]}\,.
\end{align}

\section{Dislocation Fields}
In this section, the field equation~(\ref{ME3}) will be solved.
In order to satisfy the force equilibrium condition~(\ref{FE}), 
the stress function ansatz of Mindlin-type~\citep{Mindlin63,LA09} 
should be used
\begin{align}
\label{SFA}
\sigma_{ij}=
\left(\begin{array}{cc}
\pd^2_{yy}\Phi - \pd^2_{xy}\Psi  & -\pd^2_{xy}\Phi + \pd^2_{xx}\Psi   \\\\
 -\pd^2_{xy}\Phi - \pd^2_{yy}\Psi & \pd^2_{xx}\Phi + \pd^2_{xy}\Psi  
\end{array} \right)
\end{align}
with the two stress functions $\Phi$ and $\Psi$.
It holds: $\sigma_{kk}=\Delta\Phi$ and $\sigma_{[xy]}=\frac{1}{2}\Delta\Psi$. 
A similar stress function ansatz holds for the background stress
$\sigma_{ij}^0$ in terms of the background stress functions
$\Phi^0$ and $\Psi^0$ (see Appendix A).

Substituting Eq.~(\ref{SFA}) into Eq.~(\ref{ME3}) or into Eqs.~(\ref{ME-dil})
and (\ref{ME-rot}) gives
two inhomogeneous Helmholtz equations for the stress functions
\begin{align}
\label{HE-phi}
[1 - \ell^2_1\,\Delta]\Phi &= \Phi^0\,,\\
\label{HE-psi}
[1 - \ell^2_2\,\Delta]\Psi &= \Psi^0\,,
\end{align}
where the two characteristic lengths of the 2D dislocation field theory are given by
\begin{align}
\label{l1}
\ell^2_1&=\frac{c}{2\mu(1+\nu)}\,,\\
\label{l2}
\ell^2_2&
=\frac{c}{4}\Big(\frac{1}{\mu}+\frac{1}{\gamma}\Big)
=\frac{c (\mu+\gamma)}{4\mu\gamma}\,.
\end{align}
They fulfill the relation
\begin{align}
\label{rel}
\ell_2^2=\frac{(\mu+\gamma)(1+\nu)}{2\gamma}\, \ell_1^2\,.
\end{align}
The inhomogeneous parts $\Phi^0$ and $\Psi^0$ in Eqs.~(\ref{HE-phi})
and (\ref{HE-psi}) are given by Eqs.~(\ref{phi0}) and (\ref{psi0}).

The solutions of Eqs.~(\ref{HE-phi}) and (\ref{HE-psi}) are 
(see also~\citep{LA09})
\begin{align}
\label{phi}
\Phi&= -\frac{A}{2}\,\pd_y \Big\{ r^2\ln r
+ 4\,\ell^2_1\Big[\ln r +K_0\Big(\frac{r}{\ell_1}\Big)\Big]\Big\}\,, \\
\label{psi} 
\Psi&= \frac{B}{2}\, \pd_x
\Big\{r^2 \ln r + 4\,\ell^2_2\Big[\ln r
+K_0\Big(\frac{r}{\ell_2}\Big)\Big]\Big\}\,.
\end{align}
where $K_n$ denotes the modified Bessel function of the second kind and of
order $n$ and the pre-factors are given by
\begin{align}
A=\frac{\mu(1+\nu) b}{2\pi}\,,\qquad
B=\frac{\mu\gamma b}{\pi(\mu+\gamma)}\,.
\end{align}

Substituting the stress functions~(\ref{phi}) and (\ref{psi}) 
into the stress function ansatz~(\ref{SFA}), 
the following components of the force stress tensor follow
\begin{align}
\label{Txx}
\sigma_{xx}&= -\frac{y}{r^4}\bigg\{A\Big[(y^2 + 3 x^2)
+\frac{4\,\ell^2_1}{r^2}(y^2 - 3x^2)
-2y^2\frac{r}{\ell_1}K_1\Big(\frac{r}{\ell_1}\Big)
- 2(y^2 - 3x^2)K_2\Big(\frac{r}{\ell_1}\Big)\Big]
\nonumber\\
&\quad
- B\Big[(x^2 - y^2) - \frac{4\,\ell^2_2}{r^2}(3x^2 -y^2)
+ 2x^2\frac{r}{\ell_2}K_1\Big(\frac{r}{\ell_2}\Big)
- 2(y^2 -3x^2)K_2\Big(\frac{r}{\ell_2}\Big)\Big]\bigg\}\,,\\
\label{Tyy}
\sigma_{yy}&= -\frac{y}{r^4}\bigg\{ A\Big[(y^2 - x^2)
-\frac{4\,\ell^2_1}{r^2}(y^2 - 3x^2)
-2x^2\frac{r}{\ell_1}K_1\Big(\frac{r}{\ell_1}\Big)
+ 2(y^2 -3x^2)K_2\Big(\frac{r}{\ell_1}\Big)\Big]
\nonumber\\
&\quad + B\Big[(x^2 - y^2) - \frac{4\,\ell^2_2}{r^2}(3x^2 -y^2)
+ 2x^2\frac{r}{\ell_2}K_1\Big(\frac{r}{\ell_2}\Big)
+ 2(3x^2 -y^2)K_2\Big(\frac{r}{\ell_2}\Big)\Big]\bigg\}\,,\\                             \label{Txy}
\sigma_{xy}&= \frac{x}{r^4}\bigg\{A\Big[(x^2 - y^2)
-\frac{4\,\ell^2_1}{r^2}(x^2 - 3y^2)
-2y^2\frac{r}{\ell_1}K_1\Big(\frac{r}{\ell_1}\Big)
+ 2(x^2 -3y^2)K_2\Big(\frac{r}{\ell_1}\Big)\Big]
\nonumber\\
&\quad
+ B\Big[(x^2 + 3y^2) + \frac{4\,\ell^2_2}{r^2}(x^2 -3y^2)
- 2x^2\frac{r}{\ell_2}K_1\Big(\frac{r}{\ell_2}\Big)
- 2(x^2 -3y^2)K_2\Big(\frac{r}{\ell_2}\Big)\Big]\bigg\}\,,\\
\label{Tyx}
\sigma_{yx}&=\frac{x}{r^4}\bigg\{A\Big[(x^2 - y^2)
-\frac{4\,\ell^2_1}{r^2}(x^2 - 3y^2)
-2y^2\frac{r}{\ell_1}K_1\Big(\frac{r}{\ell_1}\Big)
+ 2(x^2 -3y^2)K_2\Big(\frac{r}{\ell_1}\Big)\Big]
\nonumber\\
&\quad
- B\Big[(x^2 - y^2) - \frac{4\,\ell^2_2}{r^2}(x^2 -3y^2)
- 2y^2\frac{r}{\ell_2}K_1\Big(\frac{r}{\ell_2}\Big)
+ 2(x^2 -3y^2)K_2\Big(\frac{r}{\ell_2}\Big)\Big]\bigg\}.
\end{align}
The trace of the stress tensor is
\begin{align}
\label{Tkk}
\sigma_{kk}&= - 2 A \,\frac{y}{r^2}
\Big[1 -\frac{r}{\ell_1}K_1\Big(\frac{r}{\ell_1}\Big)\Big]\,
\end{align}
and the skew-symmetric part of the force stress tensor reads
\begin{align}
\label{Tskew}
\sigma_{[xy]}= B \,
\frac{x}{r^2}\Big[1 -\frac{r}{\ell_2}K_1\Big(\frac{r}{\ell_2}\Big)\Big].
\end{align}
It is worth noting that all the components 
of the force stress tensor~(\ref{Txx})--(\ref{Tskew}) 
are non-singular and finite everywhere 
and they agree with the force stresses of an edge dislocation
calculated in the plane strain problem using the framework of 
three-dimensional dislocation gauge theory~\citep{LA09}.
Only the pre-factor $A$ and the pre-factor of the trace of the stress tensor
are slightly different due to plane strain problem of an edge dislocation 
where $\sigma_{zz}\neq 0$ (see~\citep{LA09}).

Substituting Eqs.~(\ref{Txx})--(\ref{Tskew}) into 
Eq.~(\ref{CR-B}), the components of the elastic distortion tensor
are obtained
\begin{align}
\label{Bxx}
\beta_{xx}&=-\frac{y}{r^4}\bigg\{\frac{A}{2\mu}
\Big[(y^2 + 3 x^2)
+\frac{4\,\ell^2_1}{r^2}(y^2 - 3x^2)
-2y^2\frac{r}{\ell_1}K_1\Big(\frac{r}{\ell_1}\Big)
- 2(y^2 - 3x^2)K_2\Big(\frac{r}{\ell_1}\Big)\Big]
\nonumber\\
& \hspace{1cm}
- \frac{B}{2\mu}\Big[(x^2 - y^2) - \frac{4\,\ell^2_2}{r^2}(3x^2 -y^2)
+ 2x^2\frac{r}{\ell_2}K_1\Big(\frac{r}{\ell_2}\Big)
- 2(y^2 -3x^2)K_2\Big(\frac{r}{\ell_2}\Big)\Big]\nonumber\\
&\hspace{1cm}
-\frac{\nu b}{2\pi}\, r^2
\Big[1 -\frac{r}{\ell_1}K_1\Big(\frac{r}{\ell_1}\Big)\Big]
\bigg\}\,,\\
\label{Byy}
\beta_{yy}&=-\frac{y}{r^4}\bigg\{\frac{A}{2\mu}
\Big[(y^2 - x^2)
-\frac{4\,\ell^2_1}{r^2}(y^2 - 3x^2)
-2x^2\frac{r}{\ell_1}K_1\Big(\frac{r}{\ell_1}\Big)
+ 2(y^2 -3x^2)K_2\Big(\frac{r}{\ell_1}\Big)\Big]
\nonumber\\
&\hspace{1cm}
+ \frac{B}{2\mu}\Big[(x^2 - y^2) - \frac{4\,\ell^2_2}{r^2}(3x^2 -y^2)
+ 2x^2\frac{r}{\ell_2}K_1\Big(\frac{r}{\ell_2}\Big)
+ 2(3x^2 -y^2)K_2\Big(\frac{r}{\ell_2}\Big)\Big]
\nonumber\\
&\hspace{1cm}
-\frac{\nu b}{2\pi}\, r^2
\Big[1 -\frac{r}{\ell_1}K_1\Big(\frac{r}{\ell_1}\Big)\Big]\bigg\}
\,,\\
\label{Bxy}
\beta_{xy}&=\frac{x}{r^4}\bigg\{\frac{A}{2\mu}\Big[(x^2 - y^2)
-\frac{4\,\ell^2_1}{r^2}(x^2 - 3y^2)
-2y^2\frac{r}{\ell_1}K_1\Big(\frac{r}{\ell_1}\Big) + 2(x^2 -
3y^2)K_2\Big(\frac{r}{\ell_1}\Big)\Big]
\nonumber\\
&\hspace{1cm}
+ \frac{B}{2\mu}\Big[2y^2 + \frac{4\,\ell^2_2}{r^2}(x^2 -3y^2)
- (x^2 - y^2)\frac{r}{\ell_2}K_1\Big(\frac{r}{\ell_2}\Big)
- 2(x^2 -3y^2)K_2\Big(\frac{r}{\ell_2}\Big)\Big]
\nonumber\\
&\hspace{1cm}
+ \frac{B}{2\gamma}\,r^2\Big[1 -\frac{r}{\ell_2}K_1\Big(\frac{r}{\ell_2}\Big)\Big]\bigg\}\,,\\
\label{Byx}
\beta_{yx}&=\frac{x}{r^4}\bigg\{\frac{A}{2\mu}\Big[(x^2 - y^2)
-\frac{4\,\ell^2_1}{r^2}(x^2 - 3y^2)
-2y^2\frac{r}{\ell_1}K_1\Big(\frac{r}{\ell_1}\Big)
+ 2(x^2 -3y^2)K_2\Big(\frac{r}{\ell_1}\Big)\Big]
\nonumber\\
&\hspace{1cm}
+ \frac{B}{2\mu}\Big[2y^2 + \frac{4\,\ell^2_2}{r^2}(x^2 -3y^2)
- (x^2 - y^2)\frac{r}{\ell_2}K_1\Big(\frac{r}{\ell_2}\Big)
- 2(x^2-3y^2)K_2\Big(\frac{r}{\ell_2}\Big)\Big]
\nonumber\\
&\hspace{1cm}
-\frac{B}{2\gamma}\,r^2\Big[1 - \frac{r}{\ell_2}K_1\Big(\frac{r}{\ell_2}\Big)\Big] \bigg\}\,.
\end{align}
The elastic dilatation reads
\begin{align}
\label{Bkk}
\beta_{kk}=-  \frac{(1 - \nu)b }{2\pi} \,\frac{y}{r^2}\Big[1 -
\frac{r}{\ell_1}K_1\Big(\frac{r}{\ell_1}\Big)\Big]\,.
\end{align}
In addition the elastic rotation is
\begin{align}
\label{Bskew}
\beta_{[xy]}= \frac{\mu b}{2\pi(\mu+\gamma)} \,
\frac{x}{r^2}\Big[1 -\frac{r}{\ell_2}K_1\Big(\frac{r}{\ell_2}\Big)\Big]\,.
\end{align}
In Eqs.~(\ref{Bkk}) and (\ref{Bskew}) it can be seen
that $\ell_1$ and $\ell_2$ are the characteristic lengths for
the elastic dilatation and elastic rotation, respectively.
The two characteristic lengths of our model should be estimated by
using data from experiments: $\ell_1$ from the profile of the
elastic dilatation field and $\ell_2$ from the profile 
of the elastic rotation field.
The elastic dilatation field~(\ref{Bkk}) and the elastic rotation 
field~(\ref{Bskew}) are non-singular.
Their extremum values are:
$|\beta_{kk}(0,y)|\simeq 0.399 (1-\nu)b /[2\pi\ell_1]$ at 
$|y|\simeq 1.114 \ell_1$ and
$|\beta_{[xy]}(x,0)|\simeq 0.399\mu b /[2\pi(\mu+\gamma)\ell_2]$ at 
$|x|\simeq 1.114 \ell_2$.
In this way, the two material moduli $\gamma$ and $c$ may be determined.
$\ell_2$ can be determined from the position of the maximum 
of the elastic rotation, and $\gamma$ from the maximum value 
of the elastic rotation.
If $\ell_2$ and $\gamma$ are determined, $c$ can be computed from 
Eq.~(\ref{l2})
and $\ell_1$ can be obtained from Eqs.~(\ref{l1}) or (\ref{rel}).
Using the elastic rotation $\beta_{[xy]}$,
the new material parameters $\gamma$, $c$, $\ell_1$, and
$\ell_2$ may be determined. 
Thus, $\ell_1$ and $\ell_2$ may be used as fitting parameters in order to 
compare the experimental measurement with the presented theoretical model.
In such a manner, the numerical values of the characteristic lengths
may be determined from the experimental strain curves.

\begin{figure}[t]\unitlength1cm
\centerline{
\epsfig{figure=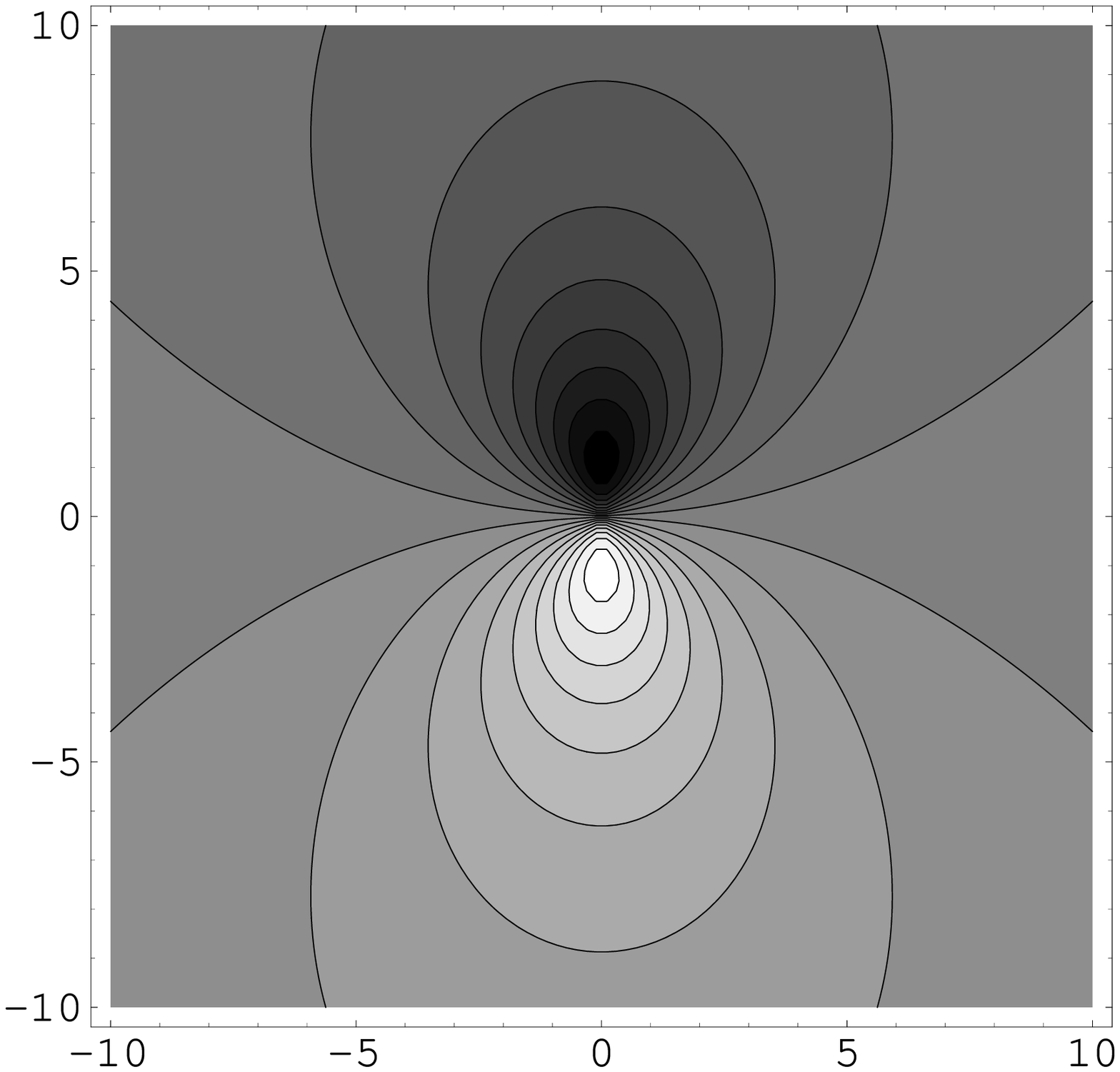,width=6.0cm}
\put(-6.7,3.0){$y/\ell_1$}
\put(-6.2,-0.3){$\text{(a)}$}
\hspace*{0.2cm}
\put(0,-0.3){$\text{(b)}$}
\epsfig{figure=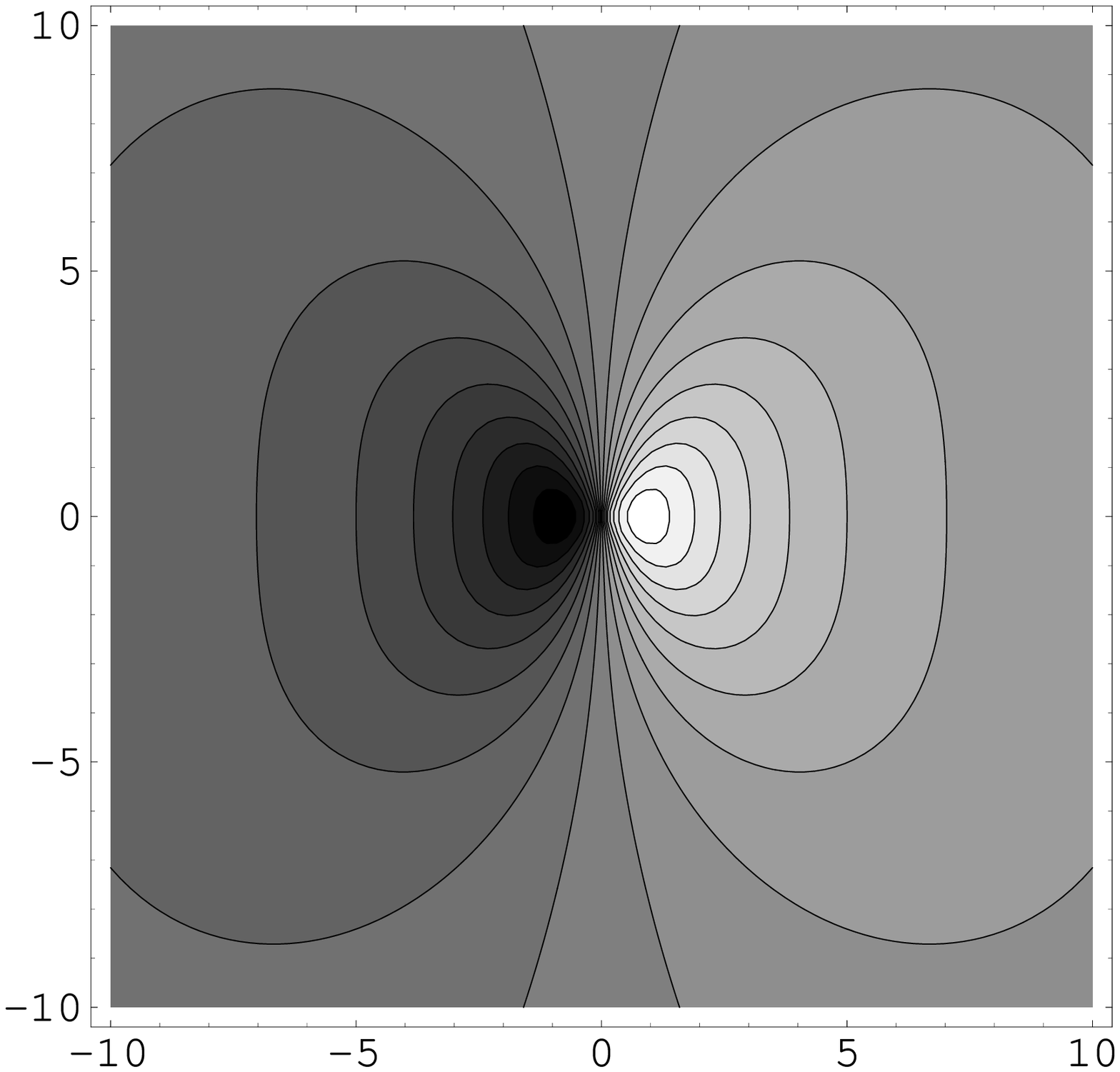,width=6.0cm}
}
\vspace*{0.2cm}
\centerline{
\epsfig{figure=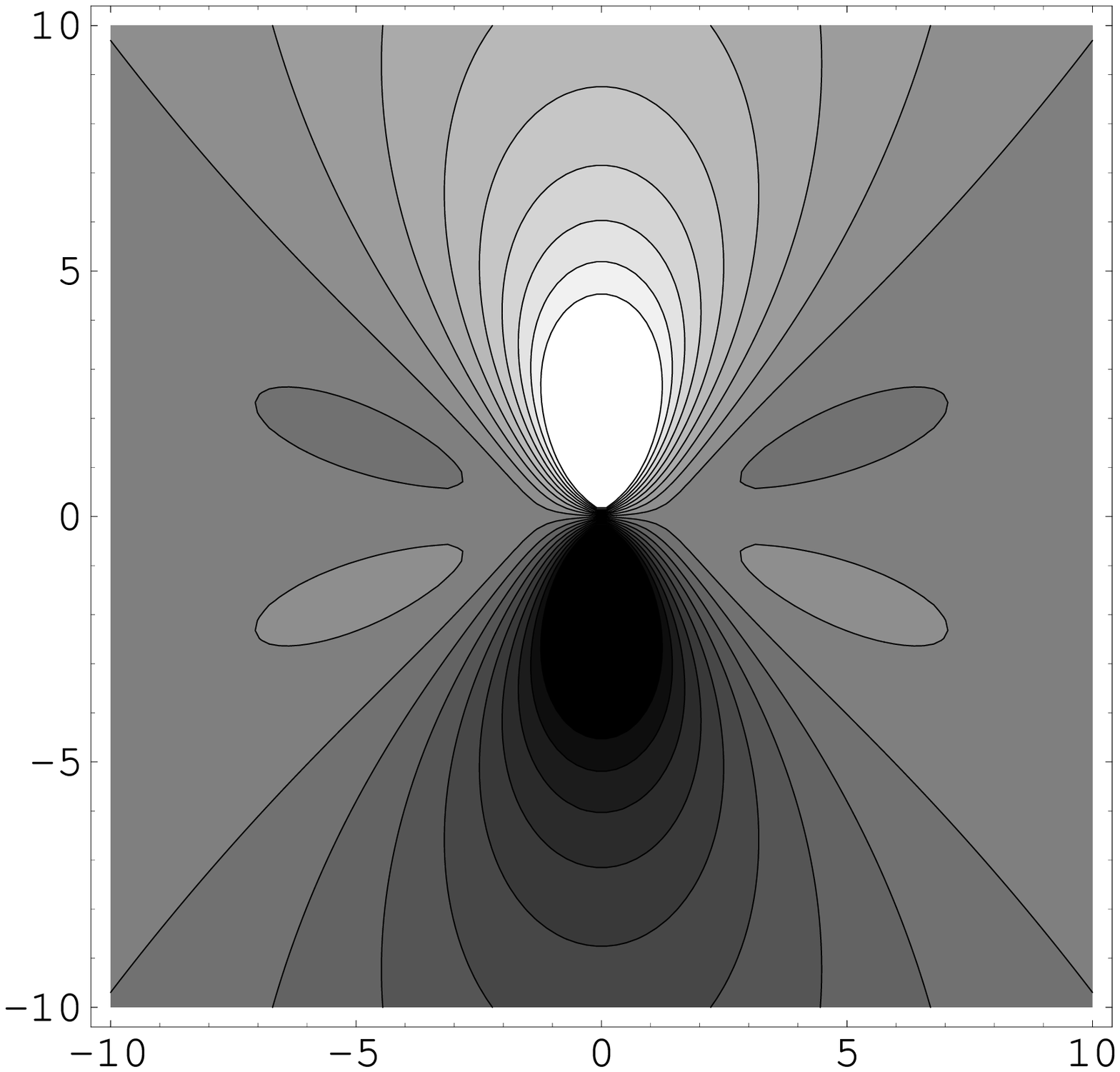,width=6.0cm}
\put(-3.0,-0.3){$x/\ell_1$}
\put(-6.7,3.0){$y/\ell_1$}
\put(-6.0,-0.3){$\text{(c)}$}
\hspace*{0.2cm}
\put(3.0,-0.3){$ x/\ell_1$}
\put(0,-0.3){$\text{(d)}$}
\epsfig{figure=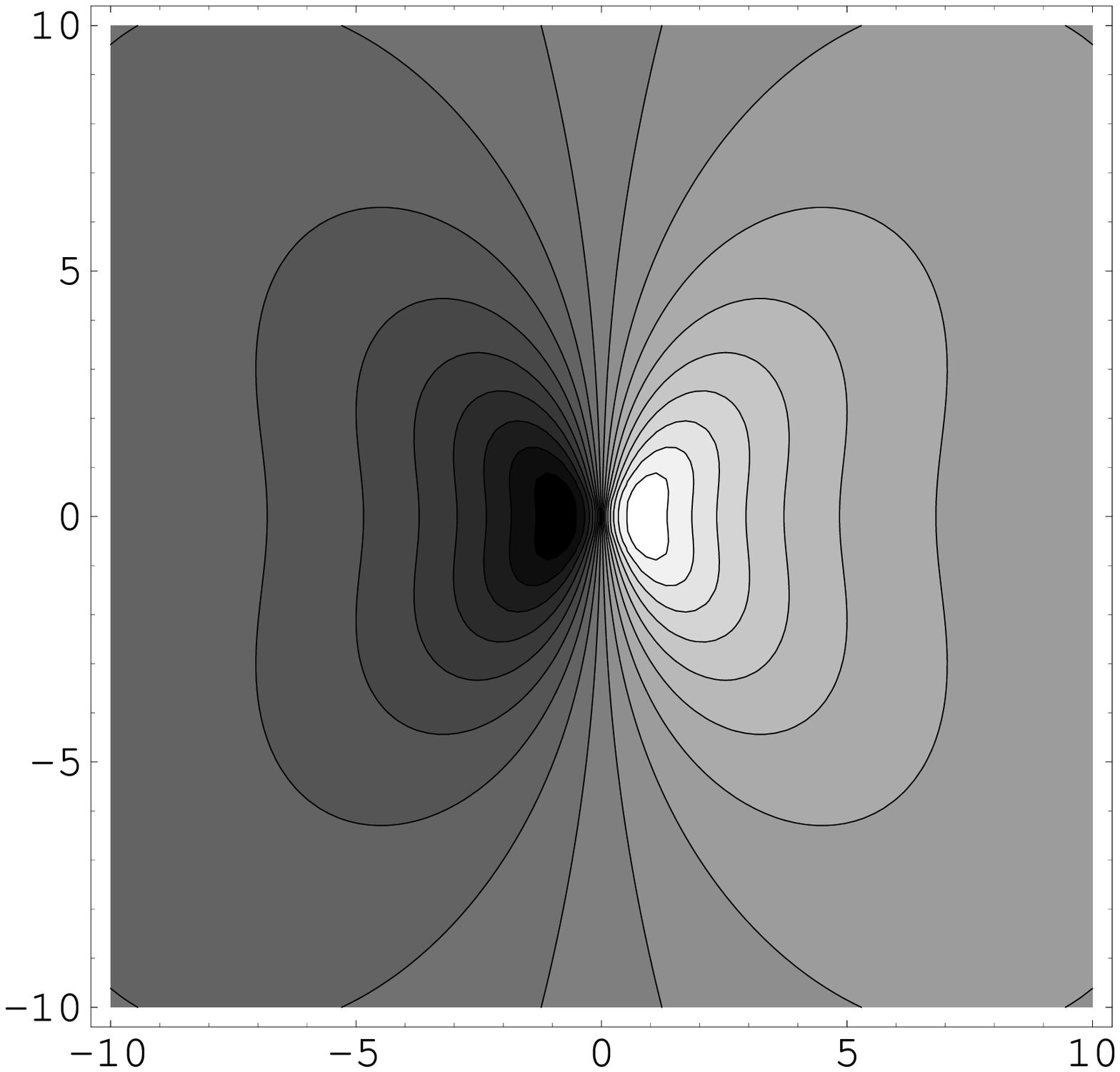,width=6.0cm}
}
\caption{Elastic distortion contours of an edge dislocation 
near the dislocation point:
(a) $\beta_{xx}$,
(b) $\beta_{xy}$,
(c) $\beta_{yy}$,
(d) $\beta_{yx}$ 
with the values: $\nu=0.12$,
 $\mu=9.95$ eV/\AA${}^2$
and $\gamma=6 \mu$.}
\label{fig:B-Co}
\end{figure}
\begin{figure}[t]\unitlength1cm
\centerline{
\epsfig{figure=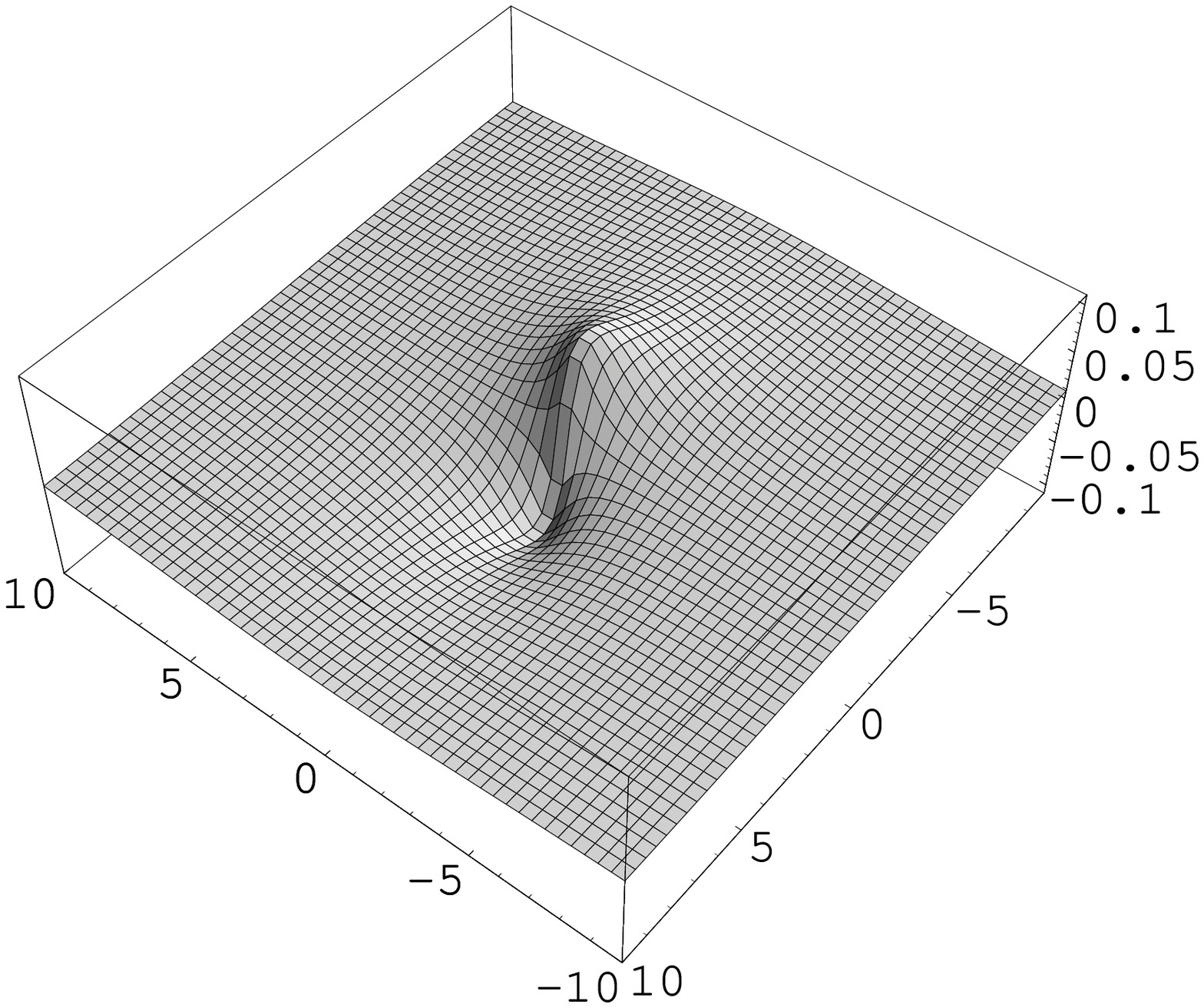,width=7.0cm}
\put(-1.5,1.0){$y/\ell_1$}
\put(-6.5,1.0){$x/\ell_1$}
\put(-6.2,-0.3){$\text{(a)}$}
\hspace*{0.4cm}
\put(0,-0.3){$\text{(b)}$}
\put(5.5,1.0){$y/\ell_1$}
\put(0.5,1.0){$x/\ell_1$}
\epsfig{figure=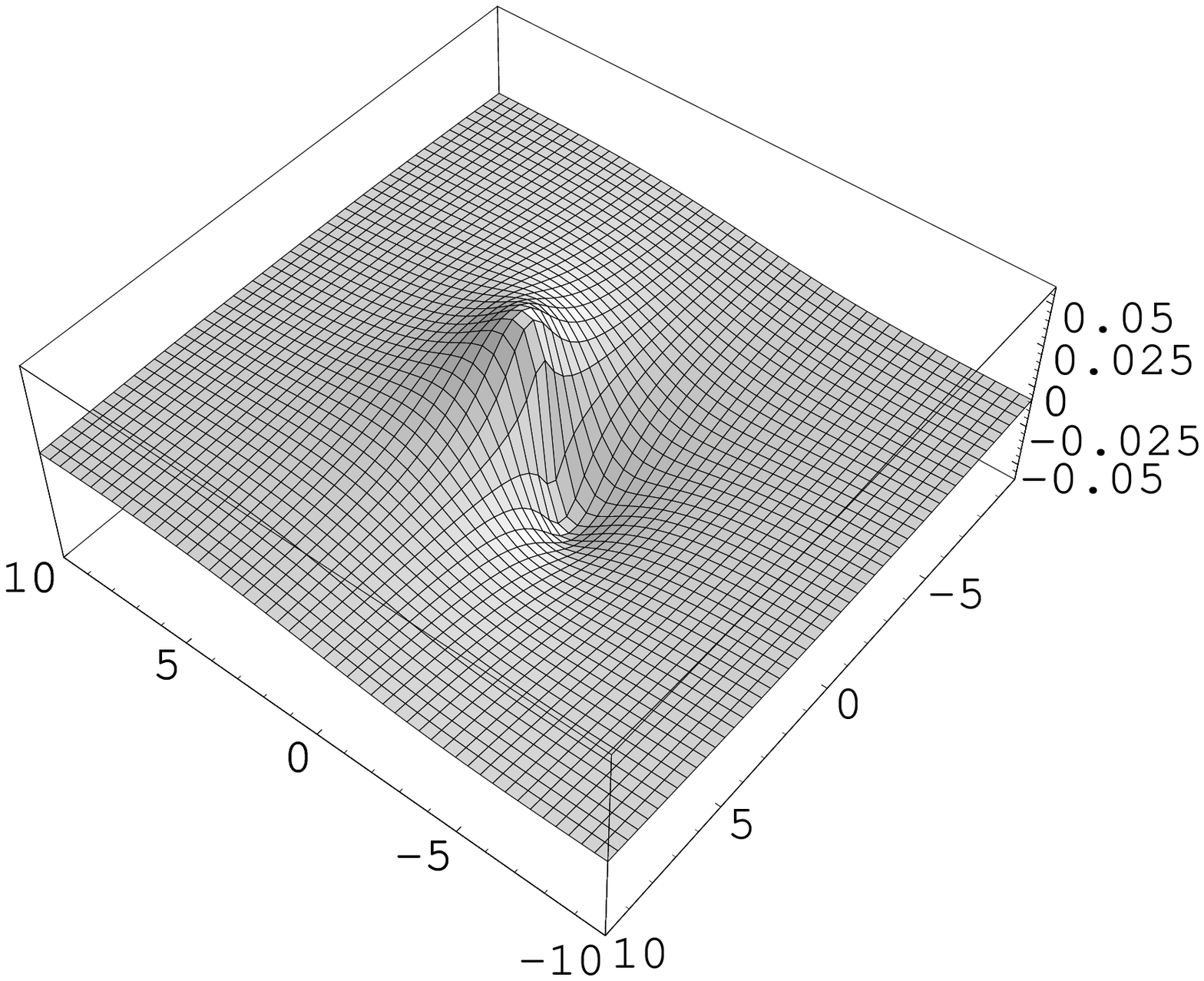,width=7.0cm}
}
\vspace*{0.2cm}
\centerline{
\epsfig{figure=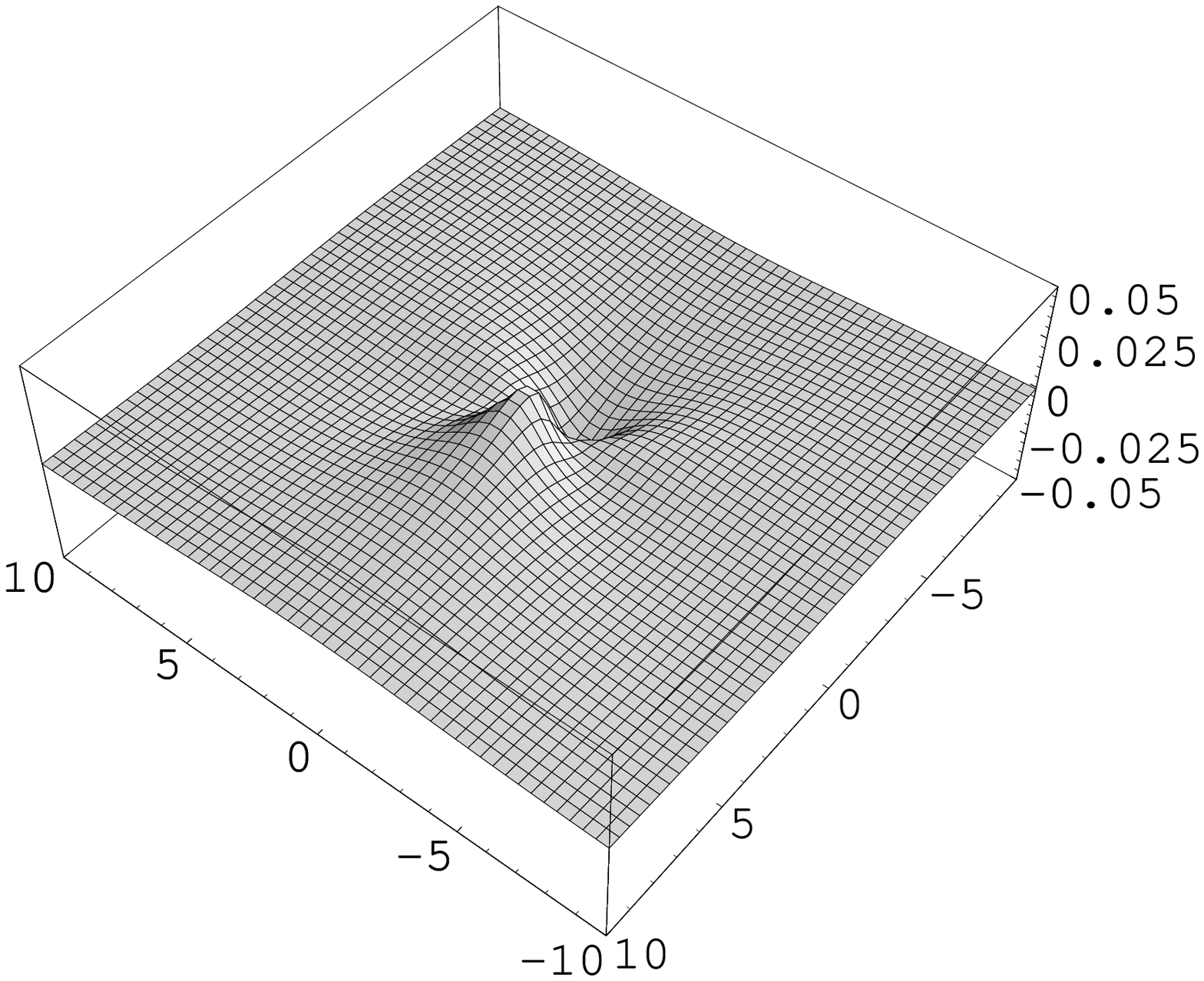,width=7.0cm}
\put(-1.5,1.0){$y/\ell_1$}
\put(-6.5,1.0){$x/\ell_1$}
\put(-6.0,-0.3){$\text{(c)}$}
\hspace*{0.4cm}
\put(5.5,1.0){$y/\ell_1$}
\put(0.5,1.0){$x/\ell_1$}
\put(0,-0.3){$\text{(d)}$}
\epsfig{figure=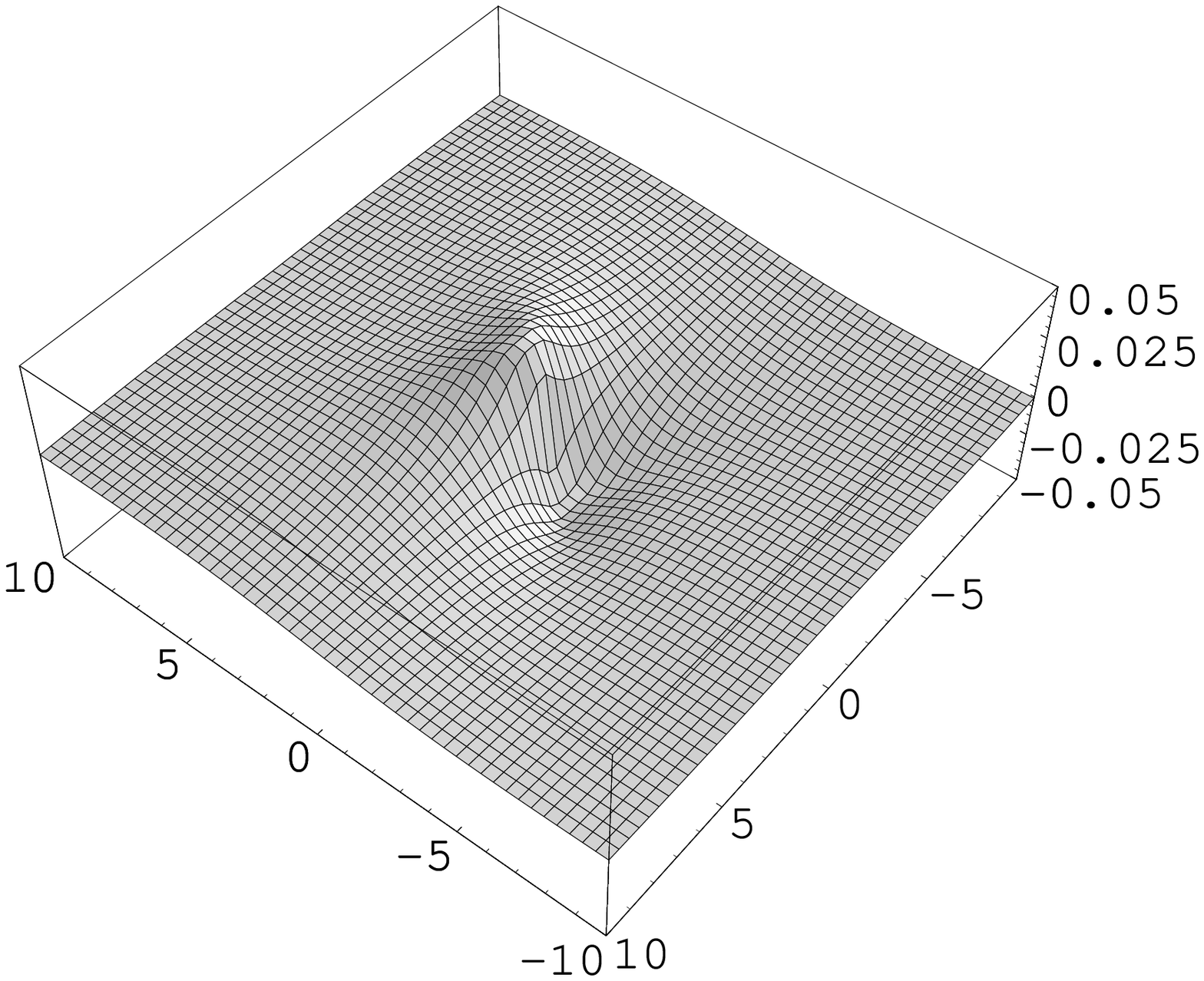,width=7.0cm}
}
\caption{Elastic distortion of an edge dislocation in units of $b$:
(a) $\beta_{xx}$,
(b) $\beta_{xy}$,
(c) $\beta_{yy}$,
(d) $\beta_{yx}$
with the values: $\nu=0.12$,
 $\mu=9.95$ eV/\AA${}^2$
and $\gamma=6 \mu$.
}
\label{fig:B-3D}
\end{figure}

The elastic distortion fields~(\ref{Bxx})--(\ref{Byx}) are plotted in
Figs.~\ref{fig:B-Co} and \ref{fig:B-3D}. 
For the plots we used the relation~(\ref{rel}) and 
the values: $\nu=0.12$, $\mu=9.95$ eV/\AA${}^2$ at 300~K~\citep{Zak}
and $\gamma=6 \mu$. 
Fig.~\ref{fig:B-Co} shows the contours of the elastic distortion fields.
The spatial distribution of the elastic distortion fields
near the dislocation point is presented in Fig.~\ref{fig:B-3D}.
Fig.~\ref{fig:B-3D} demonstrates that the elastic distortion fields are 
non-singular. Thus, there is no singularity at the dislocation point.
It can be seen in Fig.~\ref{fig:B-Co}(a) that the elastic strain $\beta_{xx}$ 
does not predict a 4-lobed strain field or strain field of butterfly shape 
as present in the classical 
isotropic elastic dislocation theory~\citep{HL}. 
Moreover, it shows exactly the shape as measured recently in~\citep{Warner}
(compare Fig.~\ref{fig:B-Co}(a) in the present paper
with Figs.~S12(h) and S13 in the supplementary materials of~\citep{Warner}).
In general, the elastic distortion fields 
have no artificial singularities in the core region  
and extremum values occur at a short distance away from 
the dislocation point (see Fig.~\ref{fig:B-3D}).

\begin{figure}[t]\unitlength1cm
\centerline{
\epsfig{figure=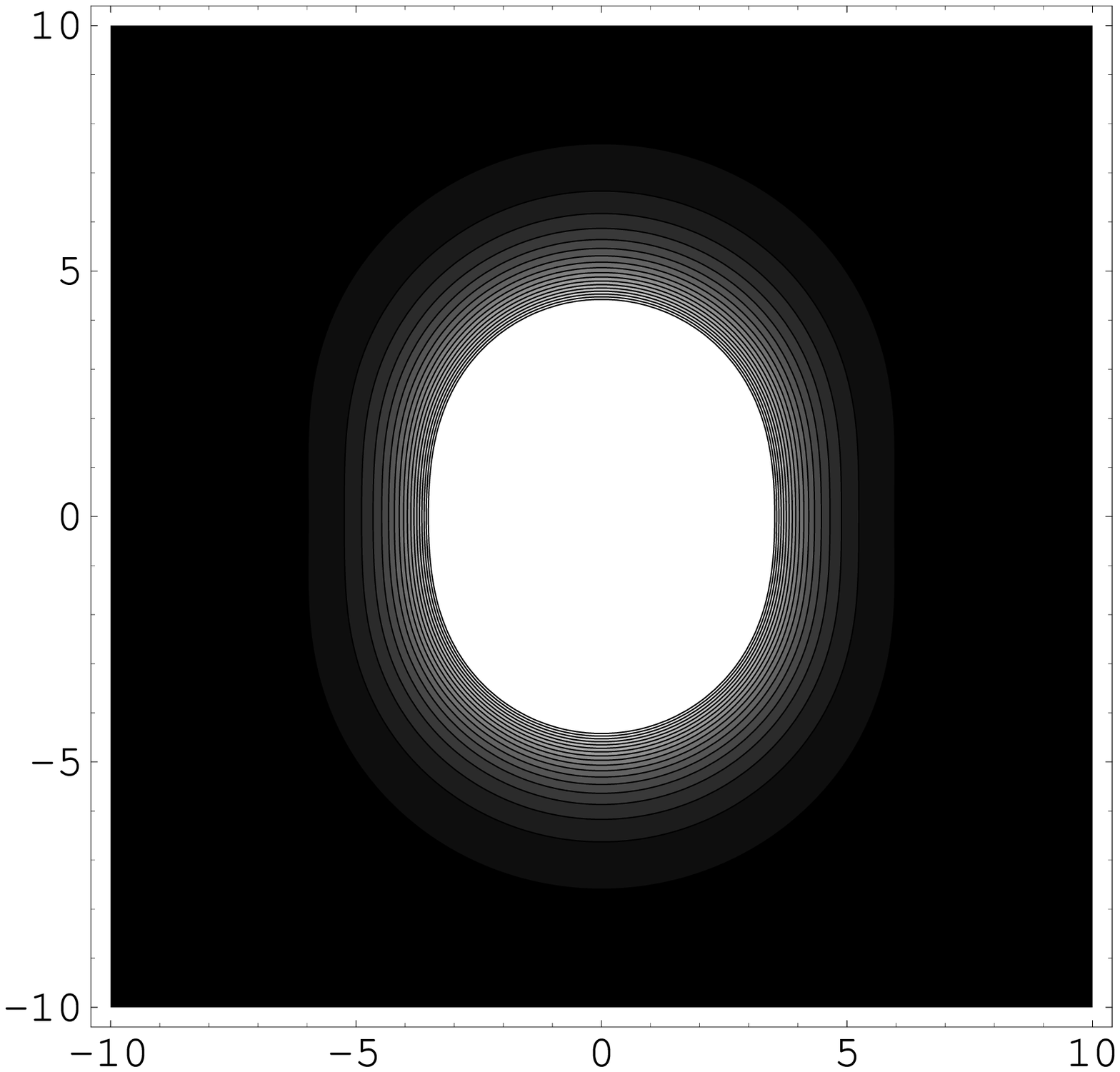,width=6.0cm}
\put(-6.7,3.0){$y/\ell_1$}
\put(-3.0,-0.3){$x/\ell_1$}
\put(-6.2,-0.3){$\text{(a)}$}
\hspace*{0.2cm}
\put(0,-0.3){$\text{(b)}$}
\epsfig{figure=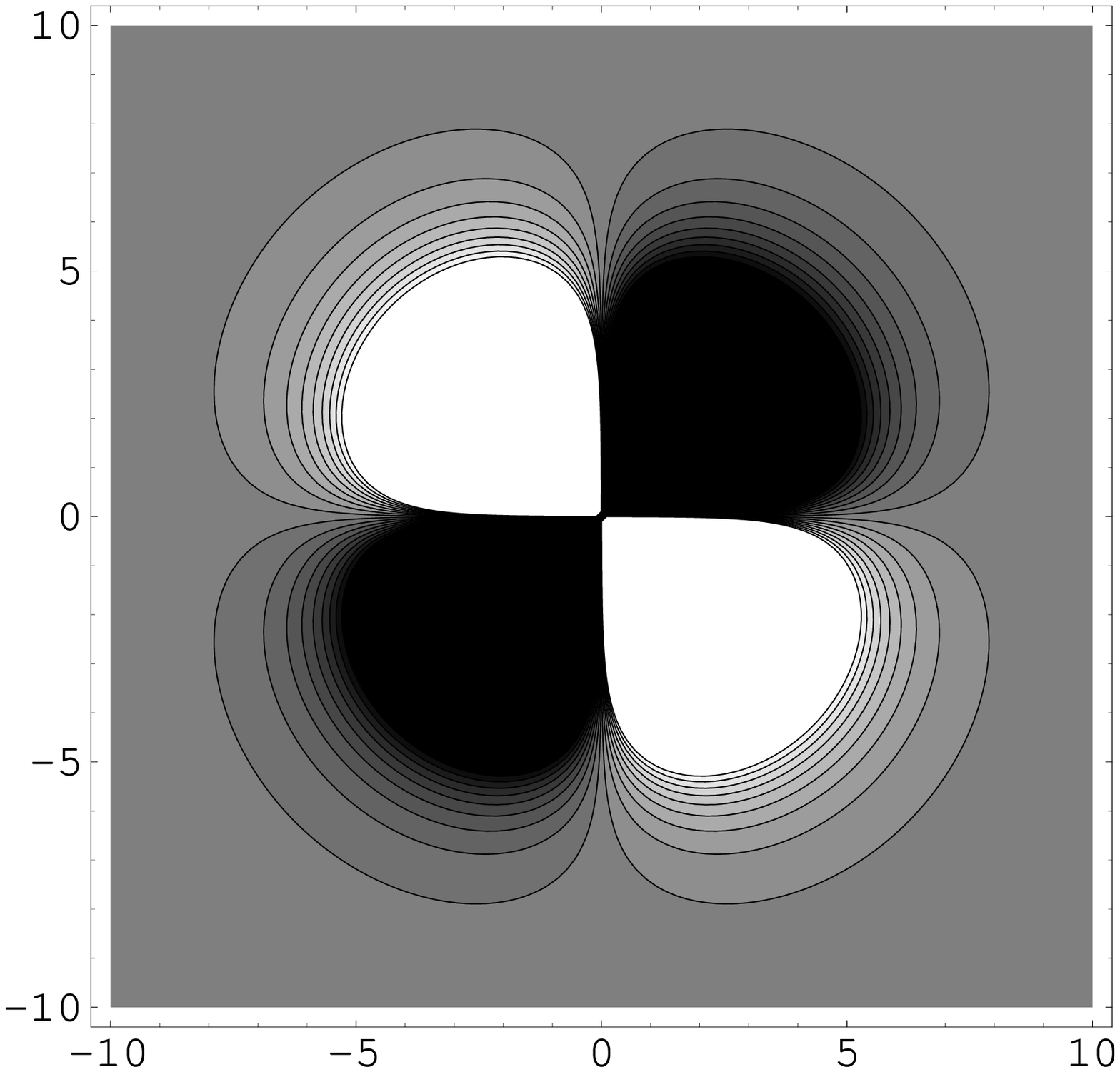,width=6.0cm}
\put(-3.0,-0.3){$ x/\ell_1$}
}
\caption{Contour plots of the dislocation density vector of 
an edge dislocation near the dislocation point:
(a) $\alpha_{x}$, (b) $\alpha_{y}$ 
with the values: $\nu=0.12$,
 $\mu=9.95$ eV/\AA${}^2$
and $\gamma=6 \mu$ (in units 
of $b/[4\pi]$).}
\label{fig:DD}
\end{figure}

Using the elastic distortion~(\ref{Bxx})--(\ref{Byx}) 
in terms of the stress functions $\Phi$ and $\Psi$, we obtain 
for the dislocation density vector of an edge dislocation 
\begin{align}
\label{DD-x}
\alpha_{x}&=-\frac{1}{2\mu(1+\nu)}\, \pd_y \Delta \Phi
+\frac{\mu+\gamma}{4\mu\gamma}\, \pd_x \Delta \Psi\,,\\
\label{DD-y}
\alpha_{y}&=\frac{1}{2\mu(1+\nu)}\, \pd_x \Delta \Phi
+\frac{\mu+\gamma}{4\mu\gamma}\, \pd_y \Delta \Psi\,.
\end{align}
Differentiating and using the Eqs.~(\ref{HE-phi}) and (\ref{HE-psi}),
the non-vanishing expressions are obtained
\begin{align}
\label{DD-x2}
\alpha_{x}&=\frac{b}{4\pi}\bigg\{
\frac{1}{\ell^2_1}\,K_0\Big(\frac{r}{\ell_1}\Big) +
\frac{1}{\ell^2_2}\,K_0\Big(\frac{r}{\ell_2}\Big)
- \frac{x^2-y^2}{r^2}
\Big[\frac{1}{\ell^2_1}\,K_2\Big(\frac{r}{\ell_1}\Big) -
\frac{1}{\ell^2_2}\,K_2\Big(\frac{r}{\ell_2}\Big)\Big]\bigg\}\,, \\
\label{DD-y2}
\alpha_{y}&=-\frac{b}{2\pi}
\frac{x y}{r^2}
\Big[\frac{1}{\ell^2_1}\,K_2\Big(\frac{r}{\ell_1}\Big) -
\frac{1}{\ell^2_2}\,K_2\Big(\frac{r}{\ell_2}\Big)\Big]\,.
\end{align}
It is worth noting that the component~(\ref{DD-y2}),
which is usually the dislocation density of an edge dislocation with Burgers
vector $b_y$, is non-zero. 
The components~(\ref{DD-x2}) and (\ref{DD-y2}) are necessary to fulfill
the dislocation equilibrium condition~(\ref{ME-SF2}).
It has been noted that these non-vanishing components 
of the dislocation density
vector do not possess cylindrical symmetry due to the $K_2$-terms 
(see Fig.~\ref{fig:DD}).
Since an edge dislocation is lacking cylindrical symmetry around the
dislocation point two length scales, $\ell_1$ and $\ell_2$, 
are needed for a proper modelling of the dislocation core of 
an edge dislocation.

\begin{figure}[t]\unitlength1cm
\vspace*{-0.5cm}
\centerline{
\begin{picture}(8,6)
\put(0.0,0.2){\epsfig{file=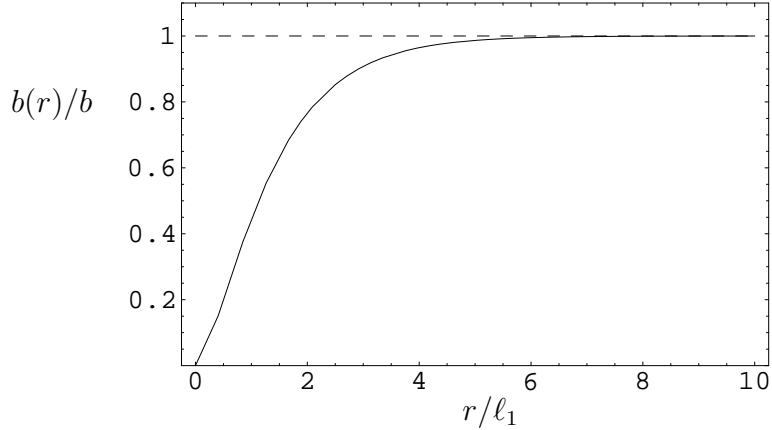,width=9cm}}
\put(4.5,-0.1){$r/\ell_1$}
\put(-1.5,4.0){$b(r)/b$}
\end{picture}
}
\caption{The modified Burgers vector of an edge dislocation $b(r)/b$
with the values: $\nu=0.12$,
 $\mu=9.95$ eV/\AA${}^2$
and $\gamma=6 \mu$ (solid).}
\label{fig:Burger-edge}
\end{figure}

Using the components~(\ref{DD-x2}) and (\ref{DD-y2})
of the dislocation density vector, the Burgers vector is calculated as
\begin{align}
\label{Burgers-x}
b(r)&=\oint (\beta_{xx}\, d x + \beta_{xy}\, d y) = \int^{2\pi}_0\int^r_0
\alpha_{x}(r',\phi')\,r'\,d r'\, d\phi'
\nonumber\\
&= b\Big\{1-\frac{1}{2}\,\Big[\frac{r}{\ell_1}
\,K_1\Big(\frac{r}{\ell_1}\Big) + \frac{r}{\ell_2}
\,K_1\Big(\frac{r}{\ell_2}\Big)\Big]\Big\}\,,\\
\label{Burgers-y}
0&=\oint (\beta_{yx}\, d x + \beta_{yy}\, d y) 
= \int^{2\pi}_0\int^r_0 \alpha_{y}(r',\phi')\,r'\, d r'\, d\phi' \,.
\end{align}
Thus, it can be seen that the 
dislocation density~(\ref{DD-y2}) does not contribute
to the Burgers vector. 
Only the $K_0$-terms in (\ref{DD-x2}) give a contribution
to the Burgers vector~(\ref{Burgers-x}).
The effective Burgers vector~(\ref{Burgers-x}) is plotted in
Fig.~\ref{fig:Burger-edge}. 
In Fig.~\ref{fig:Burger-edge}, it can be seen that 
the effective Burgers vector $b(r)$ differs from the constant Burgers
vector $b$ in the region from $r=0$ up to $r=6\ell_1$.

Last but not least,
we have to mention that all the expressions for the
elastic distortion tensor, dislocation density vector, and the effective
Burgers vector of an edge dislocation in graphene given 
in~\citep{BC12} are mistaken. The corresponding expressions 
calculated in this paper are the correct ones.

\section{Conclusions}
In this paper
we have developed a systematic dislocation continuum theory in 2D.
We have used this theory for dislocations in graphene.
We have calculated the stress and elastic distortion fields,
which are non-singular due to a straightforward regularization.
The calculated contour plots of the elastic distortion tensor
agree well with experimental data~\citep{Warner}.
The theory contains four material moduli and two characteristic 
length scales. The obtained results are useful for nano-mechanics of
2D materials (e.g., graphene).
The results are especially
important for the study of dislocations
in monolayer graphene or other 2D materials.

\section*{Acknowledgement}
The author gratefully acknowledges the grants from the 
Deutsche Forschungsgemeinschaft (Grant Nos. La1974/2-1, La1974/3-1). 

\begin{appendix}

\setcounter{equation}{0}
\renewcommand{\theequation}{\thesection.\arabic{equation}}
\section{Appendix: Edge dislocation in 2D asymmetric elasticity}
\label{appendixB}

In the case of edge dislocations, the equations of incompatibility in 2D
take the form
\begin{align}
\label{}
\alpha^0_{x}&=\beta^0_{xy,x} - \beta^0_{xx,y}\,,\\
\label{}
\alpha^0_{y}&=\beta^0_{yy,x} - \beta^0_{yx,y}\,.
\end{align}
We use the following combinations~\cite{Nowacki86}
\begin{align}
\label{A1}
A_1:=&\,\alpha^0_{y,x} - \alpha^0_{x,y} = \beta^0_{yy,xx} + \beta^0_{xx,yy} -
\beta^0_{xy,xy} -\beta^0_{yx,xy}\,,\\
\label{A2}
A_2:=& - \alpha^0_{x,x} - \alpha^0_{y,y} = \beta^0_{yx,yy} - \beta^0_{xy,xx} + \beta^0_{xx,xy} - \beta^0_{yy,xy}\, .
\end{align}
Expressing the elastic distortions in terms of force stresses, we obtain
\begin{align}
\label{A1-S}
A_1=&\frac{1}{2\mu}\,\Big(\sigma^0_{yy,xx}+\sigma^0_{xx,yy}
-(\sigma^0_{xy,xy}+\sigma^0_{yx,xy})-\frac{\nu}{1+\nu}\, 
\Delta(\sigma^0_{xx} + \sigma^0_{yy})\Big)\,,\\
\label{A2-S}
A_2=& \frac{1}{2\mu}(\sigma^0_{xx,xy} - \sigma^0_{yy,xy}) +
\frac{\gamma - \mu}{4\mu\gamma}(\sigma^0_{xy,yy} - \sigma^0_{yx,xx}) +
\frac{\gamma + \mu}{4\mu\gamma}(\sigma^0_{yx,yy} - \sigma^0_{xy,xx})\,.
\end{align}

Because we deal with asymmetric force stresses 
we use a 2D stress function ansatz
given by Mindlin for couple-stress theory~\cite{Mindlin63}
\begin{align}
\label{AM0}
\sigma^0_{ij}=
\left(\begin{array}{cc}
\pd^2_{yy}\Phi^0 - \pd^2_{xy}\Psi^0  & -\pd^2_{xy}f^0 + \pd^2_{xx}\Psi^0  \\ \\
-\pd^2_{xy}\Phi^0 - \pd^2_{yy}\Psi^0 & \pd^2_{xx}f^0 + \pd^2_{xy}\Psi^0  
\end{array} \right)\,,
\end{align}
where $\Phi^0$ and $\Psi^0$ are stress functions of second order. 
The stress function ansatz~(\ref{AM0}) is the generalization of the 
stress function ansatz with the Airy stress function $\Phi^0$ for symmetric
stresses. If $\Psi^0$ is zero, (\ref{AM0}) reduces to the usual expression for the stresses
in terms of the Airy stress function $\Phi^0$.
Equations~(\ref{A1-S}) and (\ref{A2-S}) are reduced to the
following 2D inhomogeneous bi-harmonic equations
\begin{align}
\label{}
\Delta\Delta\,\Phi^0&= 2\mu(1+\nu)A_1\,,\\
\label{}
\Delta\Delta\,\Psi^0&= - \frac{4\mu\gamma}{\mu + \gamma} A_2\,.
\end{align}
Because we consider an edge dislocation 
located at $(x,y)=(0,0)$ and with the Burgers vector $b=b_x$,
the dislocation density vector has the form
\begin{align}
\label{}
\alpha^0_{y}=0\,,\qquad \alpha^0_{x}=b\,\delta(x)\delta(y)\,.
\end{align}
In this manner, we obtain
\begin{align}
\label{BH-f}
\Delta\Delta\,\Phi^0&= - 2\mu (1+\nu) b\,\pd_y[\delta(x)\delta(y)]\,,\\
\label{BH-psi}
\Delta\Delta\,\Psi^0&= \frac{4\mu\gamma b}{\mu + \gamma}\,\pd_x[\delta(x)\delta(y)]\,.
\end{align}
Since the 2D Green function of the bi-harmonic equation
is
\begin{align}
\label{}
\Delta\Delta\,G&= \delta(x)\delta(y)\,,\qquad G= \frac{1}{8\pi}\,r^2\ln r
\end{align}
the solutions of (\ref{BH-f}) and (\ref{BH-psi}) are the following
Airy stress functions~\citep{Kroener81}
\begin{align}
\label{phi0}
\Phi^0&= - \frac{\mu (1+\nu) b}{4\pi}\, \pd_y (r^2 \ln r)\\
\label{psi0}
\Psi^0&= \frac{\mu\gamma\,b}{2\pi(\mu + \gamma)}\,\pd_x(r^2 \ln r)\,.
\end{align}
(\ref{phi0}) is the well-known Airy stress function for an edge
dislocation in 2D with Burgers vector $b_x$.
(\ref{psi0}) looks like
an Airy stress function for an edge dislocation with Burgers
vector $b_y$ with a different pre-factor.

\end{appendix}

\end{document}